\documentclass[a4paper,11pt]{article}
\usepackage{pos}
\usepackage{subcaption}

\title{Search for UHE photon sources using direction-time clustering algorithms with a BDT particle discriminant}

\author*[1]{Chaitanya Priyadarshi}
\author[1]{Jaros\l{}aw Stasielak}
\author[1]{Nataliia Borodai}
\author[1]{Megha Mogarkar}
\author[1]{Jan P\k{e}kala}
\author[1]{Dariusz G\'{o}ra}
\author[2]{Marcus Niechciol}

\affiliation[1]{Institute of Nuclear Physics Polish Academy of Sciences,\\
  PL-31-342 Krak\'{o}w, Poland}
\affiliation[2]{Center for Particle Physics Siegen,\\
University of Siegen, Siegen, Germany}

\emailAdd{chaitanya.priyadarshi@ifj.edu.pl}
\emailAdd{jaroslaw.stasielak@ifj.edu.pl}
\emailAdd{nataliia.borodai@ifj.edu.pl}
\emailAdd{megha.mogarkar@ifj.edu.pl}
\emailAdd{jan.pekala@ifj.edu.pl}
\emailAdd{dariusz.gora@ifj.edu.pl}
\emailAdd{niechciol@physik.uni-siegen.de}

\abstract{
Direction-time clustering methods for neutral particle searches are motivated by the possibility that ultra-high-energy (UHE) neutral particles are produced during flaring periods of various astrophysical sources. In this scenario, clusters of events correlated in both arrival direction and time would be generated, which, when observed, would indicate a neutral-particle flare, and a non-observation would provide upper limits on fluxes for point-like sources. Focusing on UHE photon searches, an improvement of the sensitivity of direction-time clustering methods by the application of different photon tags – such as the depth of the extensive air shower maximum ($X_{\textrm{max}}$), the lateral distribution of the total signal across an array of detectors ($S_b$), or a cumulative signal $\textit{trace-based}$ discriminant variable ($C_t$) – is investigated. A Boosted Decision Tree (BDT) is used to combine multiple observables, by which photon tagging is optimized, and photon/hadron separation is improved. The number of events required to identify a photon flare is thereby further reduced compared to using individual photon tags, across a range of flare durations. The possibility of extending the known and assumed “PeVatron” photon sources into the UHE regime by applying these clustering methods to existing and future observatory data is also discussed. The discovery of UHE neutral particles would mark the next milestone in multimessenger astrophysics, by which a new window onto the non-thermal universe would be opened.
}

\FullConference{29th European Cosmic Ray Symposium (ECRS2026)\\
29 June - 3 July 2026\\
Santiago de Compostela, Spain\\}

\begin{document}
\maketitle

\section{Introduction}
Ultra-high-energy cosmic rays (UHECRs; $E > 10^{17} eV$) are particles with the highest measured energies; however, associating them with any known astrophysical objects or events has yet to yield positive results\cite{auger_17}. If some of the UHECRs measured from Earth originated from some astrophysical flaring events (point sources), we expect them to be clustered with directional and temporal correlations. Detecting such direction-time clustering will be more probable for UHE neutral particles, like photons, as they will not be deflected by cosmic magnetic fields on their path to Earth. Although no astrophysical photons have been measured with energy above $10^{16} eV$, the recent advancements in astronomical detection techniques have rekindled the motivation for developing more sensitive methods for identifying such clustered events.

In this paper, we present an algorithm for finding direction-time clusters in UHECR data \cite{uhe_icrc25, ecrs24}, using an unbinned likelihood method. The technique was first developed to investigate correlations between neutrino events recorded by the IceCube Neutrino Observatory \cite{braun_2008, braun_2010}. We have incorporated the technique to work with cosmic ray air-shower data and strengthen particle discrimination by including distinct observables, referred to as photon tags. These tags help us differentiate between signal (photon-initiated air showers) and background (hadron-initiated air showers) events, based on different characterizations of the measured air shower events. 
The technique has been demonstrated to be sensitive in identifying a photon flare by requiring only a few events, spread across a wide range of flaring periods. 
We extend the application of the technique by searching for photon sources expected to accelerate photons in the UHE regime, using the public UHECR data of the Pierre Auger Collaboration \cite{auger_open}.

\section{Direction-time clustering analysis}

\begin{figure}
    \centering
    \includegraphics[width=0.5\linewidth]{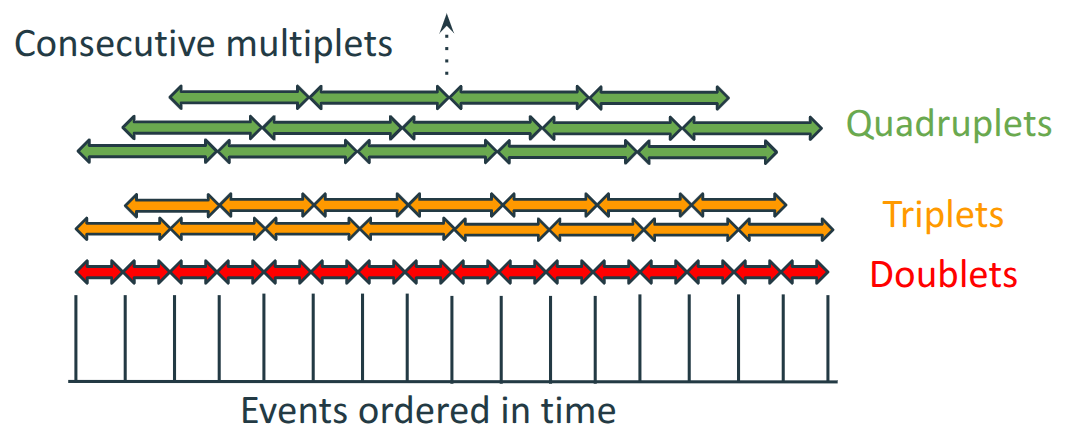}
    \captionsetup{width=0.7\linewidth}
    \caption{The Braun method includes all consecutive multiplets of events, ordered in time, for investigating the direction-time correlations.}
    \label{fig:braun_intro}
\end{figure}

The method described here is based on finding directional correlation between cosmic ray events \cite{braun_2008}, with an additional time correlation \cite{braun_2010}, by searching for all possible event multiplets, as seen in Figure \ref{fig:braun_intro}. In this method, we assume a transient astrophysical flare as the source of the clustered signal events, over an ambient background. 

For each multiplet of events $j$, we construct the likelihood $\mathcal{L}$,

\begin{equation}
\mathcal{L}(n, \Delta t_j) = \prod_{i=1}^{N} \Bigg(\frac{n}{N}s_{i} + (1 - \frac{n}{N}){b}_i\Bigg) \rm{,} \label{eq:like}
\end{equation}
where we combine the signal ($s_i$) and background ($b_i$) probability distribution functions (PDFs) for all observed events (indexed by $i$). Here, $N$ denotes the total number of events in the considered data sample, $n$ is the assumed number of signal events, and $\Delta t_j$ is the time window of the considered multiplet or doublet.  A background-only likelihood, representing the absence of signal events, simplifies to $ \mathcal{L}(0, \Delta t_j)=\prod_{i=1}^{N} b_{i}$.

For each event $i$, both the signal PDF ($s_i$) and background PDF ($b_i$) are factorized into directional, temporal, and photon tag components: 
\begin{equation}
s_i=s_i^{\mathrm{direction}}s_i^{\mathrm{time}}s_i^{\mathrm{tag}} , \hspace{1cm} b_i=b_i^{\mathrm{direction}}b_i^{\mathrm{time}}b_i^{\mathrm{tag}}, 
\end{equation}
where the photon tag components $s_i^{\mathrm{tag}}$ and $b_i^{\mathrm{tag}}$ are due to an additional discriminating factor, referred to as the photon tag, which employs PDFs to classify each event as more likely to be either a signal (photon-initiated air shower) or background (hadron-initiated event). A detailed description of these factors is postponed to Section \ref{tag_general}.
The directional component of the signal PDF ($s_i^{\mathrm{direction}}$) is modeled as a two-dimensional Gaussian distribution
$s_i^{\mathrm{direction}} = \frac{1}{2\pi \sigma_i^2} \exp \left(-\frac{\left|\vec{r}_i - \vec{r}_s \right|^2}{2\sigma_i^2}\right)$,
where $\vec{r}_i$ is the reconstructed direction of the event, $\vec{r}_s$ is the assumed source direction, and $\sigma_i$ is the angular uncertainty associated with event $i$. 
For each tested time window $\Delta t_j = t^{\rm{max}}_j - t^{\mathrm{min}}_j$, the temporal signal PDF ($s_i^{\mathrm{time}}$) is defined as 
$s_i^{\mathrm{time}}=\frac{H\left ( t_{j}^{\mathrm{max}}-t_i \right )H\left ( t_i-t_{j}^{\mathrm{min}} \right )}{\Delta t_{j}}
\label{sig_time}$,
where $H$ represents the Heaviside step function and $t_i$ is the arrival time of the event. 
This formulation ensures that only events within the specified time window $\Delta t_j$ contribute to the signal, with $s_i=0$ for all events outside this interval. 

The directional component of the background PDF is defined based on the total solid angle $\Delta \Omega$ subtended by the analyzed sky region, and the temporal component is determined by the detector uptime $\Delta T_{\mathrm{data}}$. These are based on the assumption of a uniform distribution of the background in both domains. This yields the following background PDFs: $b_i^{\mathrm{direction}} = 1 / \Delta \Omega$ and $b_i^{\mathrm{time}}=1 / \Delta T_{\mathrm{data}}$.

The corresponding test statistic ($TS$) is calculated as the logarithm of the likelihood ratio:
\begin{equation}
    TS(n, \Delta t_j) = -2 \ln \Bigg[\dfrac{\mathcal{L}(0,\Delta t_j)}{\mathcal{L}(n,\Delta t_j)} \Bigg] = 2  \sum_{i=1}^{N} \ln  \Bigg[ \frac{n}{N} \Big(\frac{s_{i}}{b_{i}} - 1 \Big) + 1\Bigg] \rm{.} \label{eq:ts_base}
\end{equation}
Using the fact that by construction, $s_i$ is non-zero only within the considered time window $\Delta t_j$, the $TS$ can be simplified to
\begin{equation}
TS(n, \Delta t_j) = 2 \sum_{i}^{N_{in,j}}\ln \Bigg(\frac{n}{N} \bigg(\frac{s_i}{b_i} - 1 \bigg) + 1 \Bigg) + 2 \cdot (N-N_{in,j})\cdot\ln \bigg(1 - \frac{n}{N}\bigg) \rm{,} \label{eq:ts_final}
\end{equation}
where the summation runs over all $N_{in,j}$ events within the $\Delta t_j$ time window. This form of the $TS$ yields approximately a tenfold improvement in computational efficiency, which is particularly important for the computationally expensive multiplet method and for calculating the statistical significance of the obtained results, which requires analysing at least $10^5$ scrambled maps.


In the application of this analysis method, we first select flare candidate events from the data based on the signal-to-background ratio of the spatial part of the PDFs, which must satisfy the following condition: 
\begin{equation}
s_i^{\rm{direction}}s_i^{\rm{tag}}/b_i^{\rm{direction}}b_i^{\rm{tag}} \geq \rm{S/B} \label{sb-ratio} \rm{,}
\end{equation}
where $i$ indexes the events and S/B is an adjustable signal-to-background threshold (note that $S/B=0$ corresponds to no pre-selection). Using these pre-selected signal-like events, we form all consecutive multiplets in the data, i.e., doublets, triplets, quadruplets, and so on. Subsequently, for each multiplet $j$ we construct the likelihood $\mathcal{L}$ and the corresponding test statistic $\rm{TS}$, taking into account all events, not only the pre-selected ones. Next, we maximize $\rm{TS}$ with respect to $n$ to obtain  $\rm{TS}_j^{\max}=\rm{TS}(n_s, \Delta t_j)$ where $n=n_s$ is the maximizing value, which varies between multiplets. As an outcome, we get a set of $\rm{TS}_j^{\max}$ values, one for each multiplet. 
The multiplet yielding the highest maximized test statistic value, i.e., $ \rm{TS}^{\rm{max}} = \rm{max} \left( \rm{TS}_j^{\max} \right)$, is then selected. The flare duration $\Delta T$ is estimated as the length of the time window of this most significant multiplet, while the corresponding $n_s$ value provides the estimate of the number of signal events, i.e., the number of direction-time correlated events.
The statistical significance of the results is assessed by comparing the highest maximized test statistic
 $\rm{TS}^{\max}$ (obtained for the most significant multiplet) 
 with a distribution of maximum test statistic values for a background, derived from many randomly generated scrambled maps mimicking the original dataset.

\section{Combination of photon tags} \label{tag_general}

Conventional photon/hadron discrimination relies on variables that exploit the physical differences between photon-initiated and hadron-initiated air showers, like the atmospheric depth of shower maximum, $X_{\rm{max}}$, or those derived from the time-integrated signal response across the array, like the shower width parameter $S_b$. One can also use the number of triggered stations, $N_{\rm{stations}}$, to see the different distributions between the two types of showers. A trace-based discriminant variable $C_t$ \cite{ct_icrc25} provides complementary information to the conventional observables. Combining the observables with methods like the Boosted Decision Tree, etc can utilize the mutual dependence of these observables and enhance the photon-hadron separation.

Monte Carlo simulations were performed to derive the values of the four observables, based on the Pierre Auger Observatory's Surface Detector array with $750m$ grid spacing (SD-750). The air showers were simulated for photons and protons, at energy $10^{17.5}$~eV using CORSIKA~7.8010 with EPOS LHC-R as the high-energy hadronic interaction model and UrQMD~1.3cr as the low-energy model. The response of the SD-750 stations was simulated and reconstructed using the Auger Offline software framework. The SD-750 array was used due to its lower energy threshold and enough real data for a statistically significant photon search. For quality cuts, we applied $X_{\rm{max}} > 450$~g/cm$^2$ (to eliminate the unreal $X_{\rm{max}}$ values, bad reconstructions). A zenith angle cut was also applied for event selection, keeping only events with zenith angles below $40^{\circ}$, to match the range in the Pierre Auger Open Data of the SD-750 array \cite{auger_open}, which will be used as the reference real dataset in this paper. The total number of simulated events after these cuts is $\sim 4.5$k photons and protons each.



An \texttt{XGBoost} \footnote{\href{https://xgboost.readthedocs.io/en/stable/}{Documentation: XGBoost}} classifier was trained on photon (signal) and proton (background) simulation data. The standard \texttt{XGBoost} tree booster, \texttt{gbtree}, was used to implement gradient boosting with 1000 decision trees, efficiently capturing non-linear relationships between the different observables. We also implemented the Out-of-Fold method to avoid over-sampling of the training and testing datasets and to find the optimal configuration for the final training of the BDT model. In addition to the four observables, Zenith was also used as an input for training the BDT model. 

\begin{figure}
    \centering
    \includegraphics[width=\linewidth]{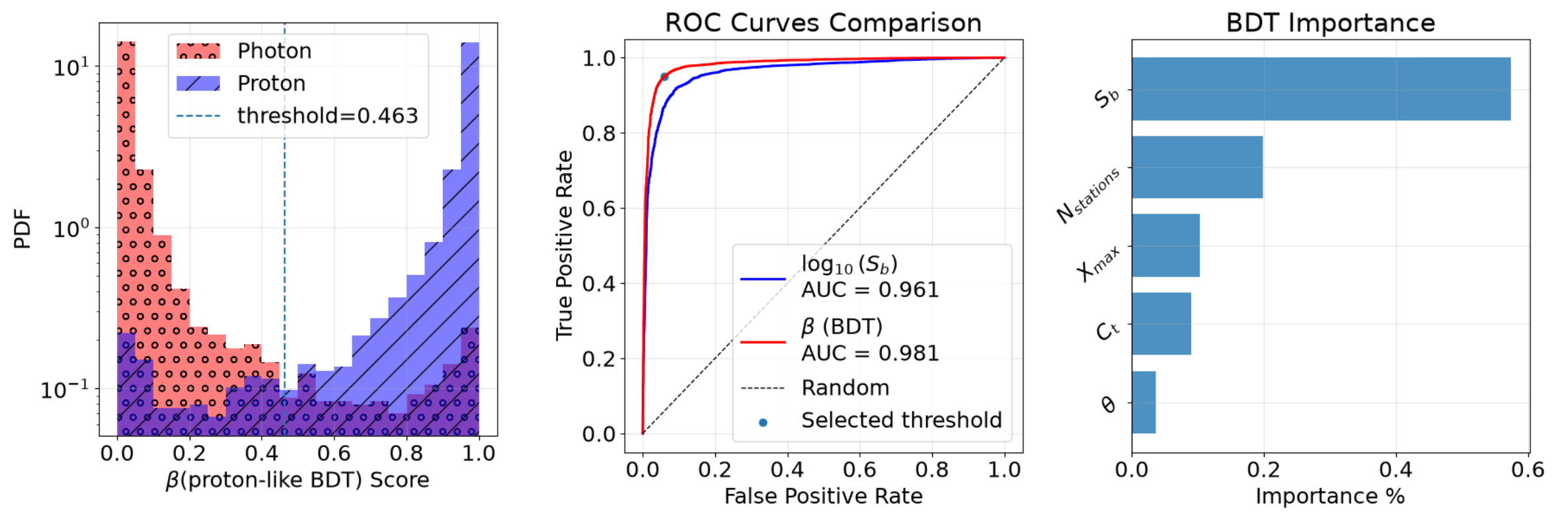}
    \caption{1. BDT distribution, 2. ROC curves, with comparison with $\log_{10}S_b$, 3. Importance of the components used to train the BDT score}
    \label{fig:bdt}
\end{figure}

The trained BDT assigns a score $\beta$ to each event in the testing data, where higher values indicate proton-like (background) events and lower values indicate photon-like (signal) events. The left panel of Figure \ref{fig:bdt} presents the PDFs of the $\beta(\rm{BDT})$ score for proton- and photon-induced showers. The PDFs show strong discrimination between the two species, which is further confirmed by the ROC (Receiver Operating Characteristic) curves shown in the central plot. 
The right plot shows the feature importance of each input component used in the training of the BDT model. The BDT score $\beta$ achieves a higher AUC (Area Under the Curve) than $\log_{10}S_b$ (earlier reference for similar studies), confirming that the multivariate combination outperforms any single observable. The PDFs of $\beta$ can thus serve as a photon tag and be incorporated into the direction-time clustering algorithm. 

\section{Monte Carlo validation test}

We investigate the sensitivity of the multiplet method based on the statistical significance of the obtained results in its application to the Auger Open Data. All events in the dataset are treated as \emph{background}, and photon flare events are simulated randomly in between them. As mentioned earlier, we compare the resulting TS distribution of the analysis with the background-only TS distribution by randomly scrambling the dataset more than $10^{5}$ times to ensure lower statistical uncertainty. The analysis is performed in a $9^\circ$ radial sky region centered on the target flaring source. 

This work was carried out using the Python package, \texttt{uhecluster}\cite{uhec, uhe_icrc25}\footnote{\href{https://uhecluster.gitlab.io/uhecluster}{Documentation: \texttt{UHECluster}}}, with the aim of developing and using different variations of the clustering algorithms (as described earlier) for UHE photon searches by existing or future Cosmic Ray observatories. The package is being developed to be accessible to any detector performing UHE neutral particle searches.

\begin{figure}
    \centering
    \includegraphics[width=0.8\linewidth]{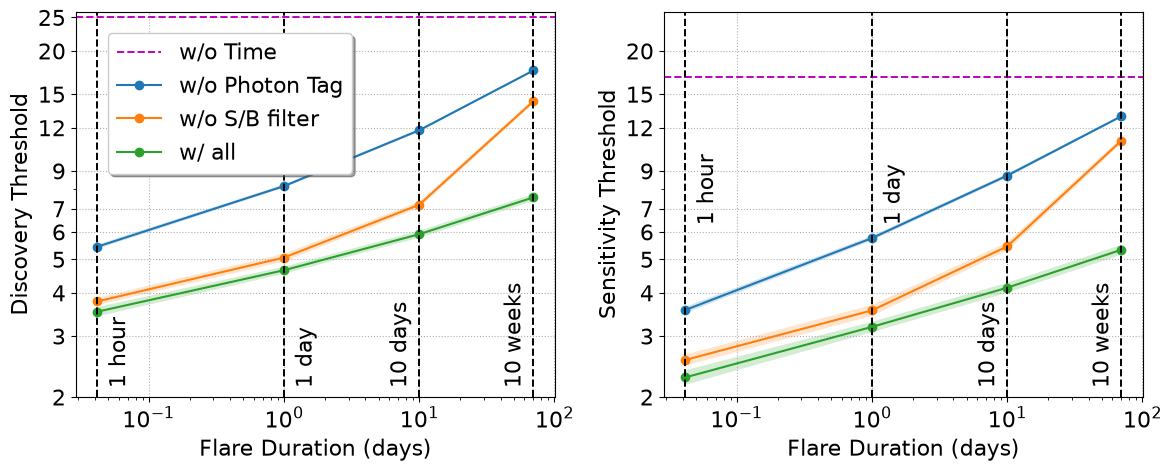}
    \captionsetup{width=0.85\linewidth}
    \caption{MC comparison of the different Braun methods, without some components, and with all}
    \label{fig:braun_comp}
\end{figure}

In this MC test, we customize the dataset by randomly assigning photon tag values, based on the PDFs derived for the photon tags for each event. Here we show the results for using the $\beta$ (BDT) tag, following the earlier established methodology\cite{uhe_icrc25}. We performed the analysis for simulated signals with flare durations of 1 hour, 1 day, 10 days, and 10 weeks, to understand the broad scope of the methods. In addition to the complete multiplets method, we also analyzed results 1) without time correlation \cite{braun_2008}, 2) without the addition of Photon Tag, and 3) without the pre-selection S/B filter. The results for the Discovery thresholds and Sensitivity thresholds can be seen in Figure \ref{fig:braun_comp}, where we see that the complete multiplet method only requires 2-5 events to reach $3\sigma$ significance, and 3-7 events to reach a $5\sigma$ significance. This is significantly stronger when compared with the other aspects of the algorithm, highlighting the overall strength of the method.

\section{Targeted search on select sources}

We now investigate the application of the multiplet method with real data to initiate a preliminary UHE photon target search. We use the Open Data for the SD-750 array\cite{auger_open} to search for the 1092 Galactic sources analyzed by Auger \cite{auger_n26} for sources of UHE neutrons. We apply the pre-selection cut, $\rm{S/B} \geq 1$, and limit the maximum flare duration search of 90 days. The photon tag values of $\beta$ are simulated for each event, assuming a priori that they are protons. The process involves 1) running the algorithm for each target source to get the reconstructed number of signal events, cluster/flare duration, and the TS for the estimation, and then comparing the estimated TS with 2) a background distribution procured by $10^5$ scrambled maps, to 3) get the p-value of the source analysis. 

\begin{figure}[htp]
\begin{subfigure}{0.5\linewidth}
    \includegraphics[width=0.9\textwidth]{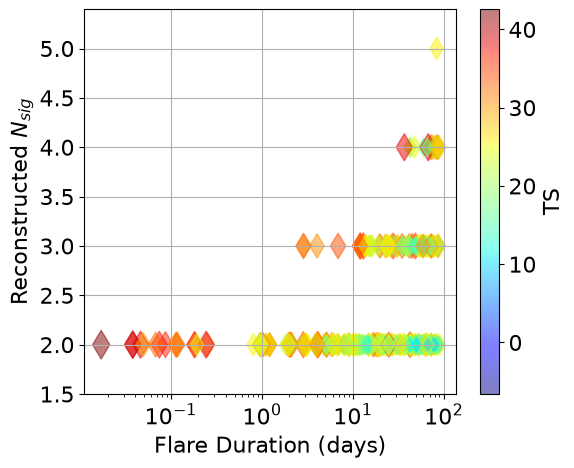}
    \captionsetup{width=0.8\textwidth}
    \caption{\textbf{Analysis Step 1:} Distribution of the reconstructed number of signal events vs reconstructed flare duration of the analyzed signal multiplet, from the first step of the analysis. The TS values are used to scale the size and color of the diamond-shaped markers.}
    \label{fig:step1}
\end{subfigure}
\begin{subfigure}{0.5\linewidth}
    \includegraphics[width=0.9\textwidth]{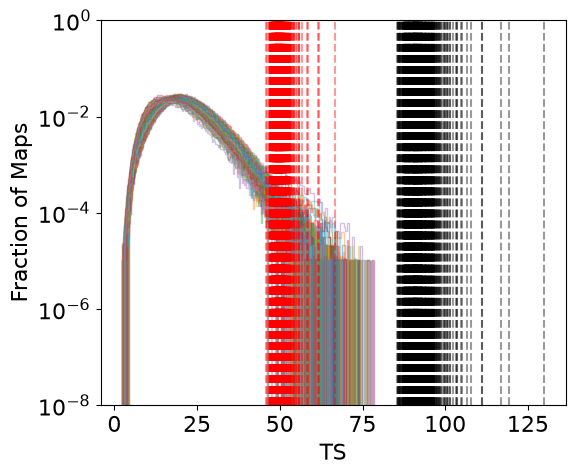}
    \captionsetup{width=0.9\textwidth}
    \caption{\textbf{Analysis Step 2:} Background TS distribution obtained by scrambling $10^5$ times the data around the target sources. The red and black dashed lines mark the $3\sigma$ and $5\sigma$ thresholds for each distribution.}
    \label{fig:step2}
\end{subfigure}
\caption{The first two analysis steps for target search}
\end{figure}

\begin{figure}[htp]
    \centering
    \includegraphics[width=0.9\linewidth]{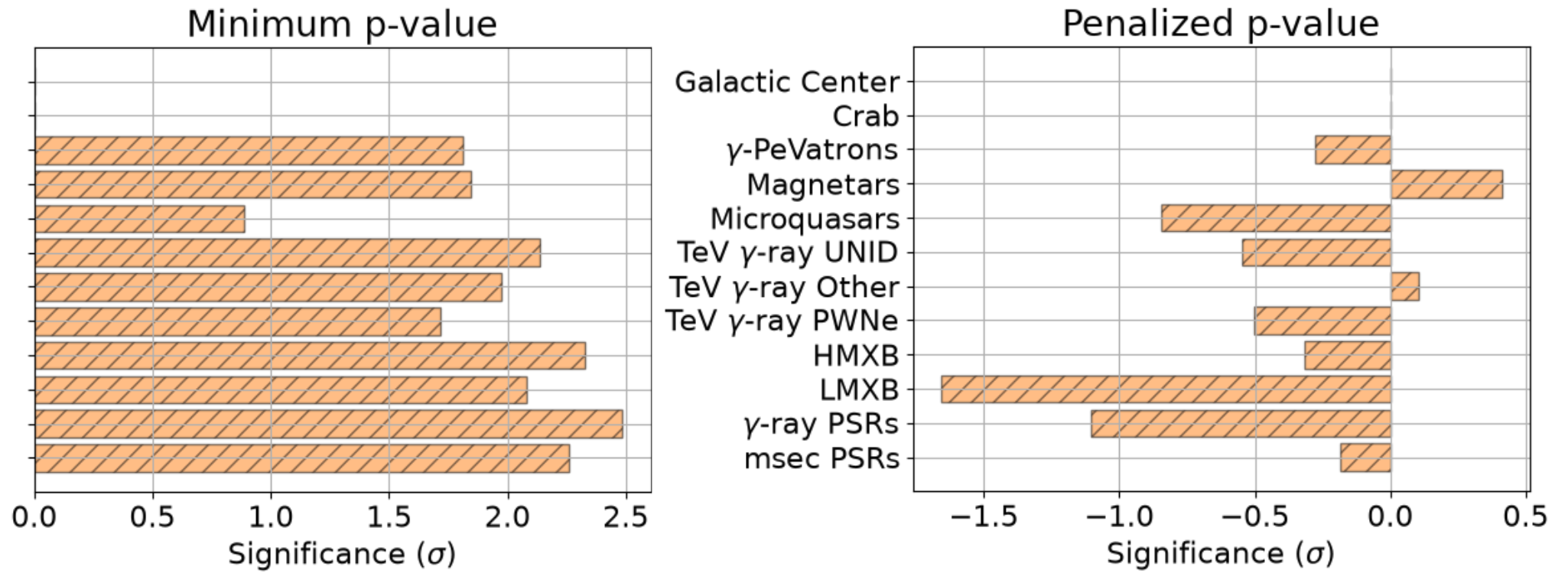}
    \captionsetup{width=0.9\linewidth}
    \caption{\textbf{Analysis Step 3:} Final p-value comparison of the different categories of sources analyzed. The left plot shows the significance of the source with the lowest p-value, and the right plot shows the significance corresponding to the penalized p-value of the respective source category.}
    \label{fig:step4}
\end{figure}

Following the first two analysis steps, some of the sources were not selected in the final analysis. Analysis of 415 sources did not yield a successful TS estimation (mostly outside the declination range of the data), and an additional 305 sources gave statistically fewer successful background reconstructions. We present the results from the analysis of the remaining 372 sources. We see a maximum of 5 reconstructed signal events across all sources. We do not see any source with more than $3\sigma$ estimated TS (see the left Figure \ref{fig:step4}), obtained from the direction-time clustering algorithm with the addition of the $\beta$ (BDT) photon tag. 

In addition to the analysis of the individual targets, we also perform stacked analysis for each category of sources \cite{auger_17}, such that we discount any relative attenuation difference from candidate sources of the same class, given an interaction length between 90 and  900 kpc of primary photons. In this, we penalize the estimated p-value for sources of each category by using,

\begin{equation}
p^* = 1 - (1 - p)^N \rm{.}
\end{equation}

This gives the probability that one or more of the $N$ candidate sources in the target set would yield a p-value smaller than $p$ by chance, assuming each p-value is independently drawn from a uniform distribution. The category-wise distribution of the significance corresponding to the penalized p-values from the results of this study, for each category of analyzed sources, is shown in Figure \ref{fig:step4}, along with those with the lowest p-values in each category.

\section{Conclusion}

In this paper, we have presented the robustness of the direction clustering method to search for short and long flares from point sources, using the Auger SD-750 array. The addition of the temporal correlation lowers the minimum number of signal events required to achieve $5\sigma$ by 3-7 times. The application of the BDT method in optimizing the signal-background discrimination is also demonstrated with the help of the clustering algorithm. Along with MC validation, we also present an example case study by applying it to the real data. The method is being further developed to better account for the detector observability constraints.

\section{Acknowledgment}
The authors would like to thank the colleagues from the Pierre Auger Collaboration for all the fruitful discussions. We want to acknowledge the support in Poland from the National Science Centre grant No. 2022/45/B/ST9/02163 and from Ministry of Science and Higher Education grant No. 2026/WK/05.


\begin{thebibliography}{99}
\bibitem{auger_17}
The Pierre Auger Collaboration (2017). \emph{A Targeted Search for Point Sources of {E}eV Photons with the Pierre Auger Observatory}. \href{https://doi.org/10.3847/2041-8213/aa61a5}{\emph{The Astrophysical Journal Letters}, 837(2):L25}

\bibitem{uhe_icrc25}
J. Stasielak, et al. (2025). \emph{Advanced techniques of searching for flares of ultra-high-energy photons from point sources}. \href{https://doi.org/10.22323/1.501.0400}{\emph{PoS}, ICRC(2025)400}.

\bibitem{ecrs24}
J. Stasielak, et al. (2024). \emph{An improved method to search for flares from point sources of ultra-high-energy photons}. \emph{Central European Astrophysical Bulletin}, 
\textsc{ArXiv ePrint}: \href{https://doi.org/10.48550/arXiv.2412.13804}{2412.13804v1}.

\bibitem{braun_2008}
J. Braun, et al. (2008). \emph{Methods for point source analysis in high energy neutrino telescopes}. \href{https://doi.org/10.1016/j.astropartphys.2008.02.007}{\emph{Astroparticle Physics}. 29(4):299-305}.

\bibitem{braun_2010}
J. Braun, et al. (2010). \emph{Time-dependent point source search methods in high energy neutrino astronomy}. \href{https://doi.org/10.1016/j.astropartphys.2010.01.005}{\emph{Astroparticle Physics}, 33(3):175-181}.

\bibitem{auger_open}
The Pierre Auger Collaboration, \emph{Auger Open Data release 2024}, \href{https://doi.org/10.5281/zenodo.10488964}{DOI:10.5281/zenodo.10488964}.

\bibitem{ct_icrc25}
N. Borodai (2025). \emph{Gamma/hadron discriminant variables in application to high-energy cosmic-ray air showers}. \href{https://doi.org/10.22323/1.501.0199}{\emph{PoS} ICRC(2025)199}.



\bibitem{uhec}
C. Priyadarshi (2026). \texttt{uhecluster (Version v0.1.0)} [Computer software]. Zenodo. \href{https://doi.org/10.5281/zenodo.22262501}{DOI: 10.5281/zenodo.22262501}

\bibitem{auger_n26}
The Pierre Auger Collaboration (2026). \emph{Search for Ultra-high-energy Neutrons from Galactic Sources with the Pierre Auger Observatory}. \href{https://doi.org/10.3847/1538-4357/ae3f05}{\emph{The Astrophysical Journal}, 999(2):168}.


\end{thebibliography}
\end{document}